\documentclass[12pt]{iopart}

\usepackage{graphicx}

\begin{document}

\title{Dynamical observation of mobility edges in one-dimensional incommensurate optical lattices}

\author{Zhihao Xu}
\address{Institute of Theoretical Physics, Shanxi University, Taiyuan 030006, China}
\address{Collaborative Innovation Center of Extreme Optics, Shanxi University, Taiyuan 030006, P.R.China}
\address{State Key Laboratory of Quantum Optics and Quantum Optics Devices,
	Institute of Opto-Electronics, Shanxi University, Taiyuan 030006, P.R.China}
\ead{xuzhihao@sxu.edu.cn}
\author{Hongli Huangfu}
\address{Institute of Theoretical Physics, Shanxi University, Taiyuan 030006, China}
\address{Collaborative Innovation Center of Extreme Optics, Shanxi University, Taiyuan 030006, P.R.China}
\author{Yunbo Zhang}
\address{Institute of Theoretical Physics, Shanxi University, Taiyuan 030006, China}
\address{Collaborative Innovation Center of Extreme Optics, Shanxi University, Taiyuan 030006, P.R.China}
\address{State Key Laboratory of Quantum Optics and Quantum Optics Devices,
	Institute of Opto-Electronics, Shanxi University, Taiyuan 030006, P.R.China}
\author{Shu Chen}
\address{Beijing National Laboratory for Condensed Matter Physics,
	Institute of Physics, Chinese Academy of Sciences, Beijing 100190, China}
\address{School of Physical Sciences, University of Chinese Academy of Sciences, Beijing, 100049, China}
\address{Yangtze River Delta Physics Research Center, Liyang, Jiangsu 213300, China}

\begin{abstract}
We investigate the wave packet dynamics for a one-dimensional incommensurate optical lattice with a special on-site potential which exhibits the mobility edge in a compactly analytic form. We calculate the density propagation, long-time survival probability and mean square displacement of the wave packet in the regime with the mobility edge and compare with the cases in extended, localized and multifractal regimes. Our numerical results indicate that the dynamics in the mobility-edge regime mix both extended and localized features which is quite different from that in the mulitfractal phase. We utilize the Loschmidt echo dynamics by choosing different eigenstates as initial states and sudden changing the parameters of the system to distinguish the phases in the presence of such system. 

\end{abstract}


\maketitle
\section{Introduction}
 More than sixty years ago, Anderson predicted and explained the well-known "Anderson localization" in his landmark paper \cite{Anderson} which has been widely recognized as one of the significant phenomena in the condensed matter. In the years since, Anderson localization has found its way across a wide range of different topics, such as electronic systems \cite{Katsumoto}, acoustic waves\cite{Strybulevych}, quantum optics \cite{Sperling,Wiersma,Aegerter,Schwartz,Lahini} and cold atomic gases \cite{Billy,Roati,Kondov,Jendrzejewski,Semeghini,Pasek,Hainaut,Richard}. A single-particle mobility edge as one of the most important concepts in a disordered system marks a critical energy $E_c$ separating localized from extended energy states and depends both on the disorder amplitudes and on the types of the disorder \cite{Mott,Evers}. In three-dimensional disordered systems, the quantum particles are free to move in the systems when the energies are above the mobility edge, whereas the energy states below $E_c$ are localized. In one- and two-dimensional cases, quantum states become localized for an arbitrary small disorder \cite{Abrahams,Ramakrishnan}. 
 
 However, the situation has changed in a one-dimensional quasi-periodic system, in which the localization and delocalization transition has drawn great attentions. One of the most famous quasi-random examples was proposed by Aubry and Andr\'{e} in 1980 \cite{Aubry}. One demonstrates that due to the self-duality characteristic \cite{Thouless}, all the eigenstates are extended or localized, which depends on the parameters of the system \cite{Sokoloff}, and there exist no mobility edges. Involved phenomena in the Aubry-Andr\'{e} (AA) model have been investigated, such as Hofstadter's butterfly \cite{Hofstadter,CRDean}, metal-insulator transition \cite{Grempel,Kohmoto,SDasSarma,Lahini1,Biddle,Biddle1,Pouranvari,Aulbach,Modugno,Larcher,Ingold}, topologically nontrivial properties \cite{Kraus,Lang,Silva,Zhihao,Zhihao1,Shiliang} and many body localization \cite{Basko,Schreiber,Iyer,Bordia}, etc.
 
 One can obtain a one-dimensional model displaying the mobility edge when the so-called self-dual symmetry is broken, such as a system with a shallow one-dimensional quasi-periodic potential \cite{Scherg,Hepeng,Diener,Ancilotto,Boers}. Another class of systems with the mobility edge by introducing a long-range hopping term \cite{Biddle1} or a special form of the on-site incommensurate potential \cite{Ganeshan} present the energy-dependent self-duality in the compactly analytic form. Recently, great attention has been paid to the properties of the intermediate phase characterized by the mobility edge in the quasi-periodic lattices, such as many-body localization in the presence of a single particle mobility edge \cite{Modak,LiXiaopeng,LiXiaopeng1,GaoXianlong,Purkayastha,Rossignolo,Zhihao2,Stellin,Kohlert} and the existence of Bose glass phase in finite temperature \cite{Errico,Barthel}. Many works have been tried to understand the relations between the energy spectral property of a disordered system and the dynamical propagation of the wave packet \cite{Kohmoto1,Ostlund,Kohmoto2,Kohmoto3,Abe,Katsanos,Geisel,Ketzmerick,Huckestein,Ketzmerick1,ZhenjunZhang,Dadras,Sinha,Santos,Laptyeva,Dalfovo,NgPRB,Lucioni,Moratti,Skipetrov,Shapiro}. A. Sinha \emph{et. al.} \cite{Sinha} study the Kibble-Zurek mechanism for generalized AA model with an energy-dependent mobility edge. Experimentally, the observation of the mobility edge has been reported in non-interacting ultra-cold atomic systems with a three-dimensional speckle disorder \cite{Kondov,Jendrzejewski,Semeghini,Pasek} and different numerical methods are proposed to estimate the position of $E_c$ \cite{Pasek,Shapiro,Delande,Fratini,Pasek1,Fratini1}. By monitoring the time evolution of the density imbalance and the global size of the atom cloud, the direct experimental research of the mobility edge in a one-dimensional quasi-random optical lattice of an initial charge-density wave state \cite{Scherg} is in good agreement with the theoretical results \cite{LiXiao}. 
 
 In this paper, we consider the wave packet dynamics in a one dimensional incommensurate optical lattice with the mobility edge in a compactly analytic form, which is described by the generalized AA model with a special form of the on-site potential. We employ the density propagation, long-time survival probability and mean square displacement to exhibit the dynamical properties of the intermediate phase and our numerical results show the dynamics of the mobility-edge regime mix both extended and localized features which is quite different from that in the multifractal phase. We also apply the Loschmidt echo dynamics to distinguish the intermediate regime from the other regimes shown in such models.

\section{Model and Hamiltonian}
 
As a concrete example, we choose a one-dimensional incommensurate optical lattice with a special form of the on-site potential, which is described by\cite{Ganeshan}
\begin{equation}\label{eq1}
\hat{H}=-J\sum_{j}(\hat{c}_j^{\dagger}\hat{c}_{j+1}+h.c.)+\sum_{j}\lambda_j\hat{n}_j,
\end{equation}
with
\begin{equation}\label{eq2}
\lambda_j=\lambda\frac{\cos{(2\pi \alpha j+\delta)}}{1-b\cos{(2\pi\alpha j+\delta)}},
\end{equation}
where $\hat{c}_j$ is the annihilation operator of the particles at $j$ site, $\hat{n}_j=\hat{c}_j^{\dagger}\hat{c}_j$ denotes the particle number operator and $J$ is the strength of the hopping term.  $\lambda_j$ is the on-site potential of a quasi-periodic form, where $\lambda$ is the strength of the chemical potential, $\alpha$ is an irrational number which is usually set as $\alpha=(\sqrt{5}-1)/2$ in the literatures, $\delta$ is an offset and  $b\in [0,1)$ is in the half open interval. When $b=0$, the system reduces to the AA model. By using self-duality characteristic, all the eigenstates are localized for $\lambda>2J$ and extended for $\lambda<2J$, while the eigenstates are multifractal at the transition point $\lambda=2J$. There are no mobility edges in the standard AA model.
For $b \ne 0$ case, the mobility edge separates the localized from extended states at energy $E_{c} =2(J-\lambda/2)/b$ \cite{Ganeshan}. 

\begin{figure}[!]
	\centerline{\includegraphics[width=12cm]{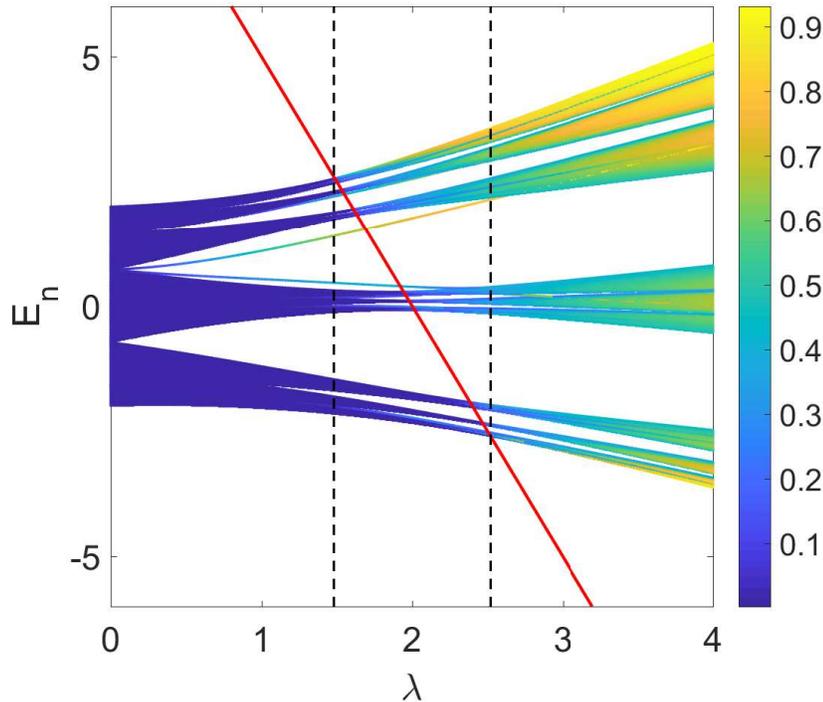}}
	\caption{IPR of the system (\ref{eq1}) with $b=0.2$, $\delta=0$ and $J$ being set as an unit energy for different $\lambda$. The red solid line shows the analytic result of the mobility edge and the black dashed lines represent the boundaries of the intermediate regime. }
	\label{Fig1}
\end{figure}

To measure the localization of the eigenstates of the system, we study the inverse participation ratio (IPR) of the eigenstate $|\psi_n\rangle$ corresponding to the eigenenergy $E_n$, $\mathrm{IPR}^{(n)}=\sum_{j}|C_j^{(n)}|^4$ \cite{Biddle1,Hepeng,Ganeshan}, containing information of the eigenstate $|\psi_n\rangle=\sum_j C_j^{(n)} |j\rangle$ with the Wannier basis $|j\rangle$ being chosen at each lattice site $j$. The IPR shows the scaling behavior with respect to the system size $L$, $\mathrm{IPR}^{(n)} \propto L^{-D_2}$ with $D_2$ being the correlation dimension of the wave function. For an extended state, $D_2=d$, where $d$ is the dimension of the system, $D_2=0$ in the localized regime and $0<D_2<1$ for a multifractal one. If there exists a value of the IPR of the energy $E_n$ which separates localized from extended states, the system exhibits a mobility edge. Fig. \ref{Fig1} shows the IPR as a function of $\lambda$ for the system (\ref{eq1}) with $b=0.2$, $\delta=0$ and $J$ being set as an unit energy. With the increase of $\lambda$, extended, intermediate and localized regimes emerge successively. The red solid line corresponds to the analytic result of the mobility edge and the intermediate regime shown in $\lambda\in (1.48,2.52)$ presents between the black dash lines.
In the next section, we will study the wave packet dynamics in the intermediate regime. As a comparison, the cases in the extended, localized and multifractal phases are also considered.

\section{Wave packet dynamics}

\begin{figure}[!]
	\centerline{\includegraphics[width=12cm]{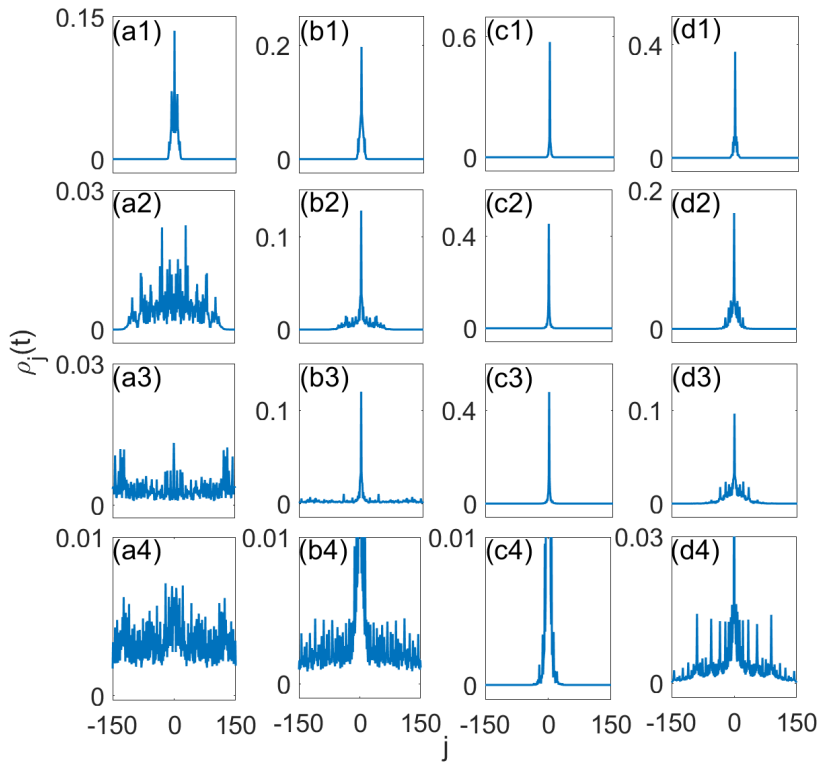}}
	\caption{Density distributions $\rho_j(t)$ for $L=301$ at different temporal times and from top to bottom rows, $t=10,100,500$ and $10^4$. Four columns correspond to different parameters of the systems: (a) $b=0.2,\lambda=1.2$, (b) $b=0.2,\lambda=1.8$, (c) $b=0.2, \lambda=2.8$ and (d) $b=0,\lambda=2$, respectively. All data in this figure are averaged $2000$ quasi-disorder realizations by choosing different phases $\delta$. }
	\label{Fig2}
\end{figure}

We investigate expansion dynamics of a wave function $|\Psi(t)\rangle$ at time $t$ governed by the Hamiltonian (\ref{eq1}). The wave function is expressed as the linear combination of the eigenstates $|\psi_n\rangle$ of the system with the corresponding eigenenergies $E_n$, the time evolution of which is accordingly
\begin{equation}\label{eq3}
|\Psi(t)\rangle=\sum_n \langle \psi_n|\Psi(0)\rangle e^{-iE_n t}|\psi_n\rangle=\sum_j C_j(t)|j\rangle,
\end{equation}
with $C_j(t)=\sum_n C_{j_0}^{(n)*}C_j^{(n)}e^{-iE_nt}$. The wave packet is initially localized at lattice $j_0$, \emph{i.e.}, $|\Psi(0)\rangle=|j_0\rangle$.

One of the important quantities we focus on is the density distribution at time $t$ given by 
\begin{equation}\label{eq4}
\rho_j(t) =|C_j(t)|^2.
\end{equation}
In Fig. \ref{Fig2}, we show the density distribution for the system with $L=301$ at different temporal times, from top to bottom rows at time $t=10,100,500$ and $10^4$, respectively. We average 2000 quasi-disorder realizations by choosing different phases $\delta$ for all the data. In the extended phase, the initial state at the center of the lattice expands rapidly and after some long-time intervals, the wave function presents a ergodic character [Fig. \ref{Fig2}(a1)-(a4)]. For $b=0.2, \lambda=2.8$, deep in the localized phase, the wave function nearly freezes its position in time which is one of signatures of the localization [Fig. \ref{Fig2}(c1)-(c4)]. In the intermediate regime, \emph{i.e.}, $b=0.2$ and $\lambda=1.8\in(1.48,2.52)$ as shown in Fig. \ref{Fig2}(b1)-(b4), the center part of the density fast decays to a finite value and the other part of the wave packet spreads similarly to that in the extended regime. For long-time dynamics, it reflects both localized and extended phenomena. The wave packet evolution in the multifractal phase is shown in Fig. \ref{Fig2}(d1)-(d4). The center part decays with time and the expanding is much slower than the one with the mobility edge. 

\begin{figure}[!]
	\centerline{\includegraphics[width=12cm]{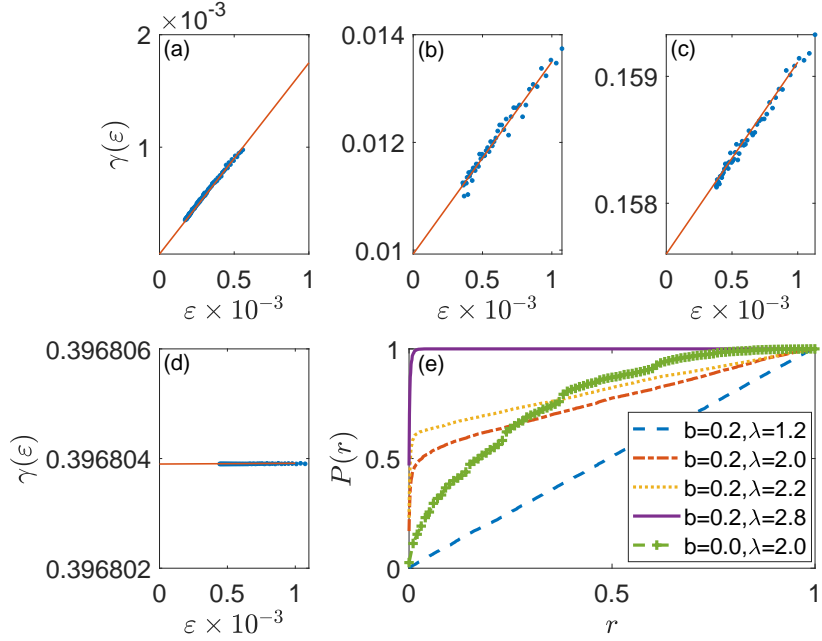}}
	\caption{Scaling  of $\gamma(\varepsilon)$ with $L=28657$, $\alpha=17711/28657$ and $b=0.2$ under periodic boundary conditions for $\lambda=1.2,2,2.2$ and $2.8$ in (a)-(d), respectively. (e) Integrated wave packet $P(r)$ at $t=10^4$ with $L=2001$ different $b$ and $\lambda$. }
	\label{Fig3}
\end{figure}

To further distinct the dynamics of the system in different phases, we observe the long-time survival probability $P(r)$ \cite{Santos}. The probability of detecting the wave packet in sites within the region $(-r/2,r/2)$ after a given time, $P(r) = \sum_{|j-j_0|\le r/2} |C_j(t\to \infty)|^{2}$, is proportional to $(r/L)^{\widetilde{D}_2}$ for finite distances where $j_0$ is at the center of the lattice and  $\widetilde{D}_2$ is the generalized dimension of the spectral measure \cite{Ketzmerick,Huckestein}. The relation of the correlation dimension $D_2$ of the wave function and $\widetilde{D}_2$ of the spectral measure is $\widetilde{D}_2=D_2/d$ for the traditional extended, localized and multifractal cases \cite{Ketzmerick,Huckestein}. For one dimensional case, $\widetilde{D}_2=D_2$. We calculate $\widetilde{D}_2$ by using the box-counting method \cite{Ketzmerick,Huckestein,Grassberger,Siebesma,YuchengEPJB,YuchengEPJB1}. Given an energy spectrum partitioned into boxes $\Omega_i(\varepsilon)$ of width $\varepsilon$ with $i\in[1,\Delta E/\varepsilon]$ and $\Delta E$ being the width of the spectrum, a quantity can be defined as
\begin{equation}\label{eq5}
\gamma(\varepsilon)=\sum_i\left[\sum_{E_n\in\Omega_i(\varepsilon)}|C_{j_0}^{(n)}|^2\right]^2\propto \varepsilon^{\widetilde{D}_2} \quad (\varepsilon\to 0),
\end{equation}
with $j_0$ being the position of the center lattice. The quantity $\gamma(\varepsilon)$ is the probability that two eigenfunctions picked from the spectral decomposition of $|\psi_n\rangle$ have an energy distance less than $\varepsilon$ \cite{Ketzmerick,Grassberger}. In Fig. \ref{Fig3}(a), we display the scaling of $\gamma(\varepsilon)$ of the spectral measures with $L=28657$ (the $23$rd number of the Fibonacci sequence) and $\alpha=17711/28657$ in different phases under periodic boundary conditions. As shown in Fig. \ref{Fig3}(a) for $b=0.2, \lambda=1.2$, $\widetilde{D}_2=1$ and $\gamma(\varepsilon)$ approximately approaches to $0$ with the decrease of $\varepsilon$ in the extended regime. In the localized regime, taking $b=0.2, \lambda=2.8$ as an example seen in Fig. \ref{Fig3}(d), $\gamma(\varepsilon)$ is finite and independence of $\varepsilon$ with $\widetilde{D}_2=0$. For $b=0.2, \lambda=2$ and $2.2$ in the intermediate regime [Fig. \ref{Fig3}(b-c)], we can see that $\widetilde{D}_2=1$ but $\gamma(\varepsilon)$ tends to a finite value when $\varepsilon \to 0$. Fig. \ref{Fig3}(e) shows the long-time survival probability $P(r)$ changes with $r/L$ at $t=10^4$ for the system with $L=2001$ in the different phases. When the parameters are in the extended regime ($b=0.2, \lambda<1.48$), since the probability of finding the wave packet at each site is the same, it linearly increases with $r$, $P(r)\propto r/L$. For the localized phase ($b=0.2, \lambda>2.52$), $P(r)$ presents exponential rise and rapidly reaches to $(r/L)^{0}=1$ \cite{Santos}. In the intermediate regime, $P(r)$ exponentially increases for $r/L\ll 1$ and for finite $r$, the increase of $P(r)$ is proportional to $r/L$ again. In contrast, for the multifractal case, \emph{i.e.} $b=0, \lambda=2$, $P(r)\propto (r/L)^{1/2}$ and $\widetilde{D}_2=1/2$ has been shown in \cite{Geisel}. The integrated wave packet $P(r)$ indicates in the regime with the mobility edge, the spreading wave packet presents mixing features of both localized and extended regimes for a long-time dynamics and is different from the multifractal dynamics. 

\begin{figure}[!]
	\centerline{\includegraphics[width=12cm]{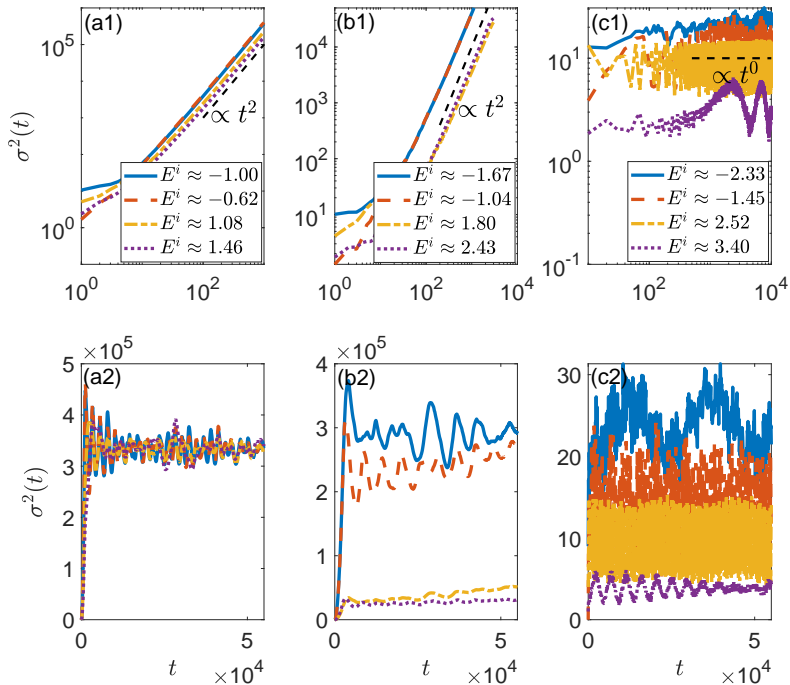}}
	\caption{Single-shot mean-square displacement $\sigma^2(t)$ as the function of time $t$ with $L=2001$, $b=0.2$, $\delta=0$ and the different initial states with selected energies. The black dashed line indicates a power-law fitting. (a1),(a2) $\lambda=1.2$; (b1),(b2) $\lambda=2$; (c1),(c2) $\lambda=2.8$. }
	\label{Fig4}
\end{figure}

\begin{figure}[!]
	\centerline{\includegraphics[width=12cm]{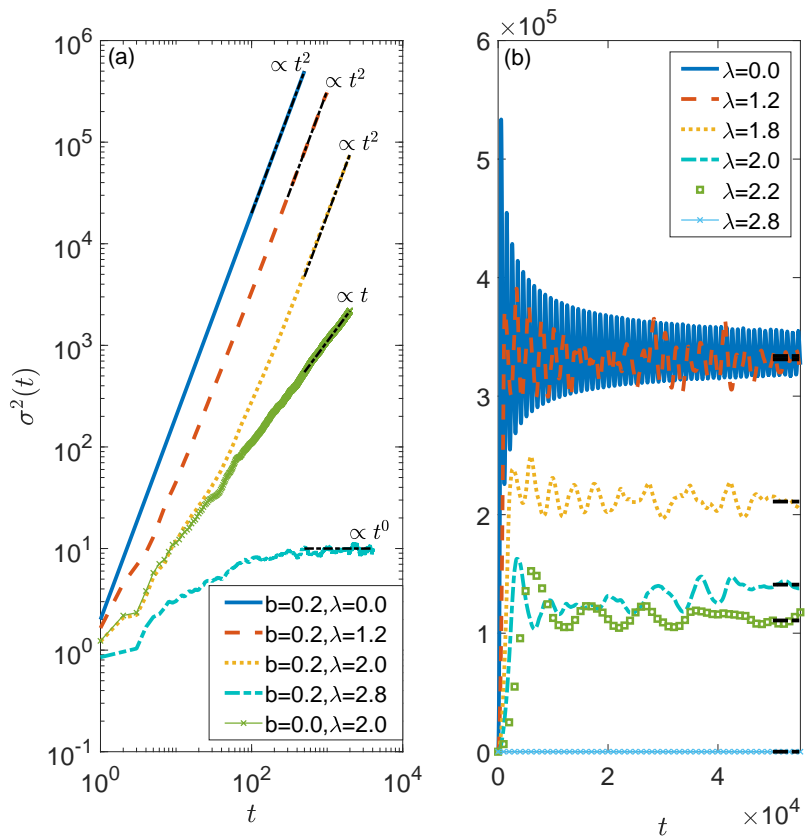}}
	\caption{(a) Log-log plot of the mean-square displacement $\sigma^2(t)$ as the function of time $t$ with averaging over the quasi-periodic configurations. The wave packet is initially localized on the center of the lattice. The black dashed line indicates a power-law fitting. (b) Time dependent of $\sigma^2(t)$ for much longer time intervals with different $\lambda$ and $b=0.2$. And the black dashed line represents the mean value of $\sigma^2(t)$ within $t\in [50000,55000]$ in steps of $10$. Here, $L=2001$ and data are averaged $100$ to $1000$ quasi-disorder realizations. }
	\label{Fig5}
\end{figure}

Mean-square displacement $\sigma^2(t)$ is an important quantity to estimate the spreading of the width of a wave packet \cite{ZhenjunZhang}, which is defined as
\begin{equation}\label{eq6}
\sigma^2(t)=\sum_j|j-j_0|^2|C_j(t)|^2.
\end{equation}
The value of $\sigma^2(t)$ grows in a power-law form of time given by $\sigma^2(t)\propto t^{\mu}$ during the expansion process. Firstly, we do not take quasi-disorder average into account. One follows the evolution of a wave packet initially localized at site $j_0$. We consider a quench protocol, where the initial wave packet at site $j_0$ is the eigenstate of Hamiltonian (\ref{eq1}) in atomic limit ($J=0$), and whose dynamics is taken with Hamiltonian (\ref{eq1}) for $J=1$ in this case. Essentially, the initial energy is fully encoded in the initial state, being precisely equal to $\lambda_{j_0}$. In Fig. \ref{Fig4}, we calculate $\sigma^2(t)$ for different initial energies $E^{i}$ with fixed $\delta=0$. For a clean system $\lambda=0$, it is clear that the mean-square displacement displays a ballistic diffusion with $\mu=2$ [see blue solid line in Fig. \ref{Fig5}(a)] and oscillates around a given value after some diffusion time intervals. We know that the wave packet expands for a clean system in the long-time limit with nearly the same probability amplitude at each site, \emph{i.e.}, $C_j(t)\sim 1/\sqrt{L}$ and the mean value of the mean-square displacement $\overline{\sigma^2}\approx 3.34\times 10^5$ for $L=2001$ which is an upper bound as shown in Fig. \ref{Fig5}(b). For the initial states with different energies at $\lambda=1.2$, the dynamical behaviors are the same as the clean case [see Fig. \ref{Fig4}(a1)] and $\overline{\sigma^2} \sim 10^5$ [see Fig. \ref{Fig4}(a2)]. For $\lambda=2.8$ where the post-quench system in the localized regime shown in Fig. \ref{Fig4}(c1),(c2), the power-law indices $\mu$ are equal to $0$ for different initial energies and $\overline{\sigma^2} \sim 10$. For the intermediate case, we take $\lambda=2$ as an example where the mobility edge at $E_{c}=0$. The mean-square displacements of both initial energies greater and smaller than the mobility edge display the ballistic diffusion with $\mu=2$ [Fig. \ref{Fig4}(b1)]. In Fig. \ref{Fig4}(b2), the mean value of the mean-square displacement for the initial state with the energy smaller than $E_{c}$ is of order $10^5$. However, for the energy of the initial state above the mobility edge, $\overline{\sigma^2}$ is much smaller than that below $E_{c}$. As shown in Fig. \ref{Fig4}(b2), for the initial energy $E^{i} \approx 1.80$, the mean value of $\sigma^2(t)$ within $t\in [50000,55000]$ in steps of $10$ amounts to $5.0855 \times 10^4$ and for $E^{i} \approx 2.43$, $\overline{\sigma^2} \approx 3.1132 \times 10^4$. To understand the results of $\overline{\sigma^2}$ in the intermediate regime, we define the probability of the projection of the initial state to the final eigenstates with the final energies above the mobility edge \cite{Shapiro}, \emph{i.e.} 
\begin{equation}\label{eq8}
P(E^{i},>)=\sum_{E^{f} > E_{c}}\left|\langle \Psi_i^{E^{i}} | \Psi_f ^{E^{f}}\rangle \right|^2,
\end{equation}
where $|\Psi_i^{E^{i}}\rangle$ is the initial eigenstate with the energy $ E^{i} $ and $| \Psi_f ^{E^{f}}\rangle $ is the final eigenstate with the energy $E^{f}$ and $P(E^{i},<)=1-P(E^{i},>)$ for the final energies below $E_c$. We take $E^{i} \approx -1.67$ and $2.43$ as examples to do our calculations for $b=0.2, \lambda=2, \delta=0$ and $L=2001$. When the initial energy is smaller than $E_c$, \emph{i.e.}, $P(-1.67,<)=0.8643$ and $P(-1.67,>)=0.1357$, the projection of the initial state to the final extended part is dominant. However, $P(2.43,<)=0.097$ and $P(2.43,>)=0.903$ for the initial energy larger than $E_{c}$ where the projection of the initial state to the extended part is much smaller than to the localized one and $\overline{\sigma^2}$ is greatly decreased. Our results shows that the power-law index $\mu$ does not depend on the choice of the initial state but depends on the post-quench regime and the mean value of the mean-square displacement $\overline{\sigma^2}$ is strongly influenced by the initial energy for the single quasi-disorder realization case in the intermediate regime and this can be of relevance for experiments with a much smaller number of realizations available.

In Fig. \ref{Fig5}, we present the mean-square displacement as the function of $t$ with $L=2001$, different $\lambda$ and $b$ and all the data are averaged $100$ to $1000$ quasi-disorder realizations. The wave packet is initially localized at the center of the lattice. As shown in Fig. \ref{Fig5}(a), the power-law increasing of the time-dependent $\sigma^2(t)$ in the extended and intermediate regimes shares the same behavior as that of the clean system. The extracted power-law indices imply that the dynamical evolution in both extended and intermediate phases is a ballistic process, in contrast to the zero power-law index for $\lambda=2.8, b=0.2$ corresponding to the localized process. We also calculate $\sigma^2(t)$ in the multifractal regime with $\lambda=2$ and $b=0$, which shows the power-law index $\mu=1$. According to our results, the power-law index of the mean-square displacement is not changed by considering the quasi-disorder average and is twice $\widetilde{D}_2$, \emph{i.e.}, $\mu=2\widetilde{D}_2$ which is in agreement with \cite{Geisel,Ketzmerick}. A theoretical analysis about the origin of the ballistic behavior is made in Ref. \cite{Leboeuf} by a Wentzel-Kramers-Brillouin semiclassical approximation. Fig. \ref{Fig5}(b) shows the distributions of the time-dependent $\sigma^2(t)$ for much longer time intervals with $L=2001$, $b=0.2$ and $\lambda=0,1.2,1.8,2,2.2$ and $2.8$, respectively. After some time intervals, the mean-square displacement oscillates around a given value and the black dashed line in Fig. \ref{Fig5}(b) represents the mean value of the mean-square displacement $\overline{\sigma^2}$ within $t\in[50000,55000]$ in steps of $10$. 
$\overline{\sigma^2}\approx 3.32\times 10^5$ for $\lambda=1.2,b=0.2$ and comparing with the clean case, the relative deviation is less than $0.6\%$. When the system is driven through the intermediate regime, $\overline{\sigma^2} \approx 2.1\times 10^5$ for $\lambda=1.8$, $\overline{\sigma^2} \approx 1.4\times 10^5$ for $\lambda=2.0$ and $\overline{\sigma^2} \approx 1.1\times 10^5$ for $\lambda=2.2$ as shown in Fig. \ref{Fig5}(b). We can see that  $\overline{\sigma^2}$ reduces with the increase of $\lambda$ due to the decreasing of extended part in the spectrum. When all the eigenstates are localized, $\overline{\sigma^2}$ is much smaller, such as $\lambda=2.8, b=0.2$ case, $\overline{\sigma^2}\approx 7.86$. It indicates that though the power-law indices of the mean-square displacement for the wave packet expanding in the extended and intermediate regimes are the same, the values of $\overline{\sigma^2}$ depend on the proportion of the localized part in the energy spectrum.


\begin{figure}[!]
	\centerline{\includegraphics[width=12cm]{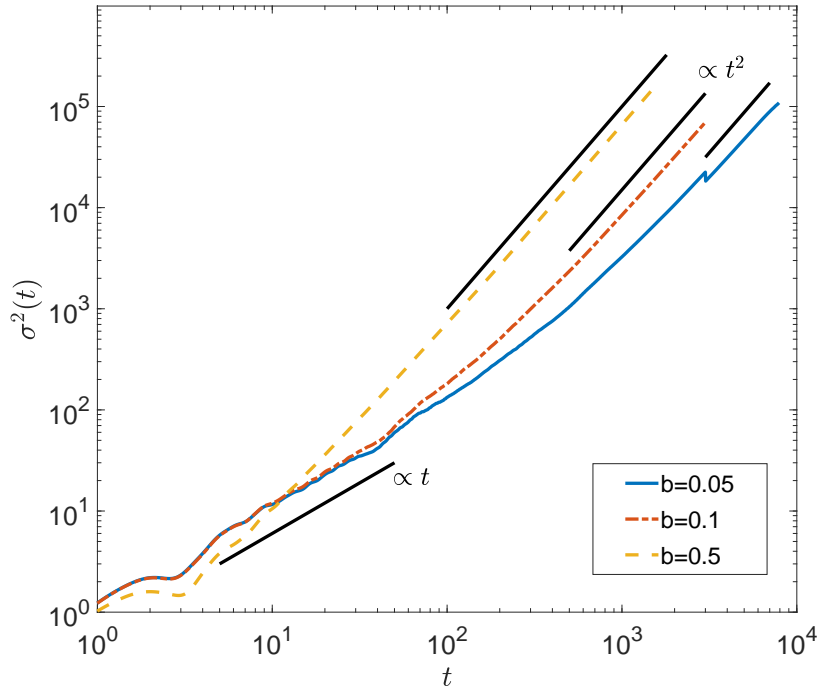}}
	\caption{Log-log plot of $\sigma^2(t)$ as the function of $t$ with different $b$, $\lambda=2$ and $L=2001$. The data are averaged $100$ different quasi-disorder realizations. }
	\label{Fig6}
\end{figure}

As mentioned above, at $b=0, \lambda=2$, the curve of the mean-square displacement as for the Hamiltonian with a sharp localized to extended phase transition has a $t$ scaling. However, for a finite $b$ and $\lambda=2$, the intermediate phase emerging, the curve has a $t^2$ scaling for long time. The case of different values of $b$ requires further exploration. Fig. \ref{Fig6} shows the log-log plot of $\sigma^2(t)$ as the function of $t$ with different $b$, $\lambda=2$ and $L=2001$. In short time, $\sigma^2(t)$ spread as $t$ for small $b$ cases, \emph{i.e.} $b=0.05$ and $0.1$ shown in Fig. \ref{Fig6}, since time scale is not enough to distinguish the energy scale defined by $b$ and $\lambda=2$. With the increase of $b$, the transition time of $t$ scale decreases. As shown in Fig. \ref{Fig6}, it hardly detects such region at $b=0.5$. For longer time, $\sigma^2(t)$ deviates away from $t$ and becomes $t^2$ since eventually the extended part dominates. 

\section{Loschmidt echo dynamics}

\begin{figure}[!]
	\centerline{\includegraphics[width=12cm]{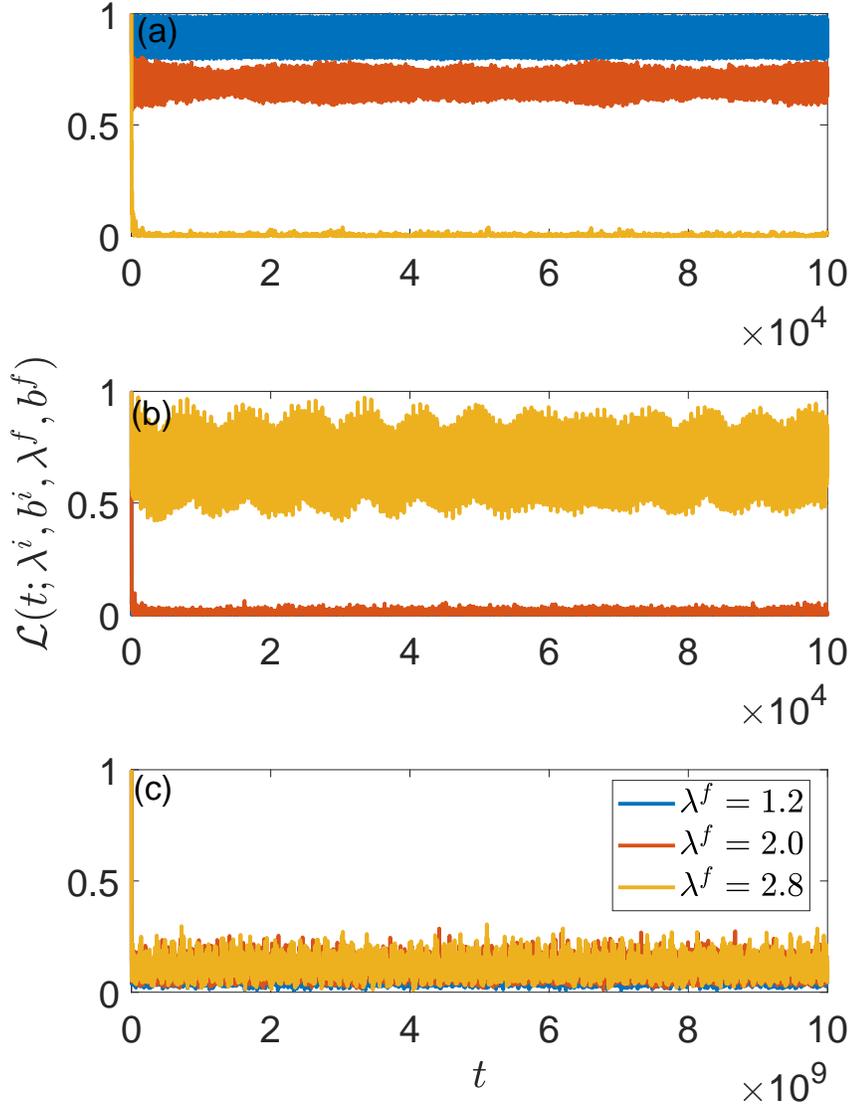}}
	\caption{Evolutions of Loschmidt echo $\mathcal{L}(t;\lambda^{i},b^{i},\lambda^{f},b^{f})$ with $b^{f}=0.2$ and different $\lambda^{f}$. The initial states are fixed as the ground states of the Hamiltonian with (a) $\lambda^{i}=0$, (b) $\lambda^{i}\to \infty$ and (c) $\lambda^{i}=2, b^{i}=0$. Here, (a),(b) $L=2001$ and (c) $L=12001$. }
	\label{Fig7}
\end{figure}

Loschmidt echo is a powerful method for analyzing nonequilibrium dynamics \cite{Heyl,Karrasch,Canovi,Jalabert,Cucchietti,Gorin,HTQuan,Jafari,Budich,Hickey,Vajna,Vajna1,Sharma,Bhattacharya,YangChao,Kennes,Szpak}, which can exhibit a series of zero points if the initial and the post-quench systems are located in different phases at some time intervals. Up to now, it has been successfully applied in a series of models, such as transverse field Ising model \cite{Heyl}, XY model \cite{Hickey,Vajna}, topological models \cite{Budich,Vajna1,Sharma,Bhattacharya}, Hubbard and Falicov-Kimball models \cite{Canovi} and disorder models \cite{YangChao}. C. Yang \emph{et al.} \cite{YangChao} suggest that the Loschmidt echo dynamics can characterize the localization-delocalization transition in the standard AA model. If both the initial and post-quench system are in the extended regime or localized regime, the values of the Loschmidt echo are always positive and if they locate in different regimes, the oscillations of the Loschmidt echo decay to zero in some time intervals. However, the behavior of the Loschmidt echo for the system with the mobility edge is still puzzled. We in the following consider the system being initially prepared in an eigenstate of the Hamiltonian $\hat{H}(\lambda^{i},b^{i})$ and then quenched to the final Hamiltonian $\hat{H}(\lambda^{f},b^{f})$. The Loschmidt echo can be defined as
\begin{equation}\label{eq7}
\mathcal{L}(t;\lambda^{i},b^{i},\lambda^{f},b^{f}) =\left|\langle \Psi(\lambda^{i},b^{i})|e^{-it\hat{H}(\lambda^{f},b^{f})}|\Psi(\lambda^{i},b^{i})\rangle \right|^2,
\end{equation}
where $|\Psi(\lambda^{i},b^{i})\rangle$ denotes the eigenstate of the initial Hamiltonian with the  parameters $\lambda^{i}$ and $b^{i}$, and the superscript $i$ ($f$) is corresponding to before (after) the quench. Fig. \ref{Fig7} shows the evolutions of Loschmidt echo with $b^{f}=0.2$, different $\lambda^{f}$, $L=2001$ for (a)-(b) and $L=12001$ for (c). The initial state is the ground state of the system with $\lambda^{i}=0$, $\lambda^{i}\to \infty$ and $\lambda^{i}=2, b^{i}=0$ shown in Fig. \ref{Fig7}(a)-(c), respectively. We can see the oscillations of Loschmidt echo display a similar behavior, when the parameters $\lambda^{f},b^{f}$ after the quench process are located in either the extended or intermediate regime with $\lambda^{i}=0$. The evolution of Loschmidt echo without decaying for long-time intervals can not touch zeros but rapidly decays to zero for $\lambda^{i}\to \infty$, which is shown in Fig. \ref{Fig7}(a)-(b). However, for the cases quenched to the localized phase $\lambda^{f}=2.8$, the evolutions of Loschmidt echo present the opposite results compared with those in former cases. We also calculate the evolution of Loschmidt echo quenched from $\lambda^{i}=2, b^{i}=0$ (a multifractal ground state) to different regimes of the Hamiltonian (\ref{eq1}) with $b^{f}=0.2$. As shown in Fig. \ref{Fig7}(c), the Loschmidt echoes can approach zero in long-time intervals, which is consistent with conventional conclusions \cite{YangChao}.

\begin{figure}[!]
	\centerline{\includegraphics[width=12cm]{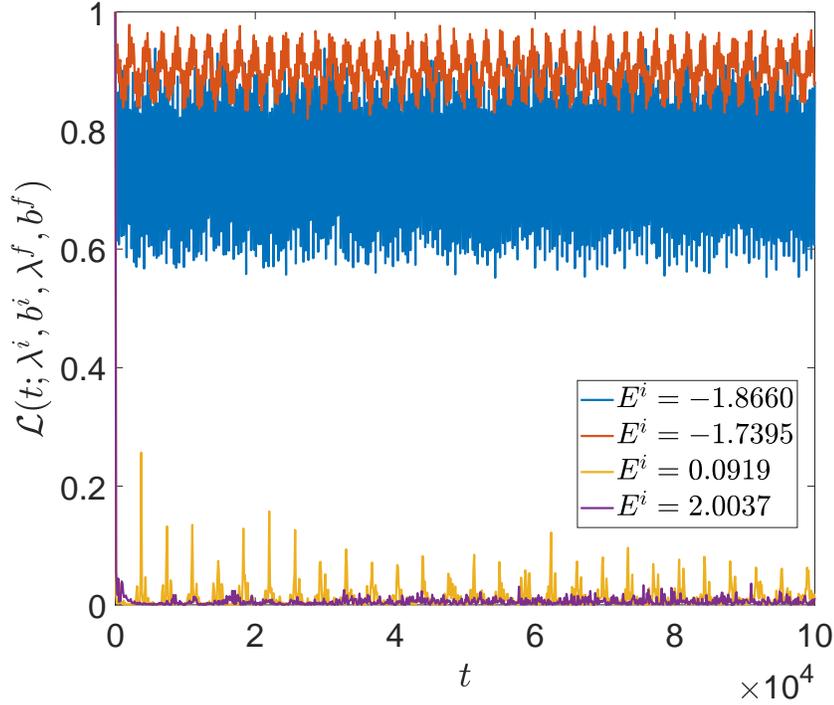}}
	\caption{Evolutions of Loschmidt echo $\mathcal{L}(t;\lambda^{i},b^{i},\lambda^{f},b^{f})$ with $\lambda^{i}=2$, $b^{i}=0.2$ and different initial eigenstates quenched to $\lambda^{f}=1.2$, $b^{f}=0.2$. Here, $\delta=0$ and  $L=2001$. }
	\label{Fig8}
\end{figure}

We notice that similar results are shown for quenched to extended and intermediate regimes from both limits ($\lambda^{i}=0$ and $\lambda^{i}\to \infty$). To further discriminate extended and intermediate regimes by Loschmidt echo dynamics, we consider the quench process from the eigenstates in the intermediate regime with different initial energies $E^{i}$ to the extended regime.  Fig. \ref{Fig8} shows the evolutions of Loschmidt echo with $\lambda^{i}=2$, $b^{i}=0.2$ and different initial eigenstates quenched to $\lambda^{f}=1.2$, $b^{f}=0.2$ located in extended regime. As mentioned above, we know that for $\lambda^{i}=2$, $b^{i}=0.2$, it is deep in the intermediate regime with the mobility edge at $E_c=0$. We choose the initial eigenstates with the energies smaller or greater than $E_c$. As shown in Fig. \ref{Fig8}, for an extended eigenstate with energy smaller than $E_c$ ($E^{i}=-1.8660$ and $E^{i}=-1.7395$ for $\delta=0$), the Loschmidt echo oscillates without decaying for long time and has a positive lower bound. If the eigenstate of intermediate regime with the energy greater than $E_c$ quenches to the extended regime ($E^{i}=0.0919$ and $E^{i}=2.0037$ for $\delta=0$), the decay of $\mathcal{L}(t;\lambda^{i},b^{i},\lambda^{f},b^{f})$ is obvious and the evolution of Loschmidt echo approaches zero after some time intervals. The results suggest that the dynamics of Loschmidt echo can distinguish the intermediate regime from the extended, localized and multifractal ones.

\section{Conclusions}

In summary, we study the spatial expansion of a wave packet in a one-dimensional incommensurate optical lattice system with a special form of on-site potential described by Eq.(\ref{eq2}). The extended, intermediate, localized and multifractal phases can be found in such system. By observing the density propagation, long-time survival probability and mean-square displacement of the wave packet in these regimes, our numerical results indicate that the dynamics of the wave packet in the intermediate phase behaves as a mixture of extended and localized phases. The evolution of Loschmidt echo is also considered to distinguish different phases emerging in such model.

\bigskip
Z. Xu is supported by the NSF of China under Grant No. 11604188 and STIP of Higher Education Institutions in Shanxi under Grant No. 2019L0097.  Y. Zhang is supported by NSF of China under Grant No. 11674201. S. Chen was supported by the NSFC (Grant No. 11974413) and the NKRDP of China (Grants No. 2016YFA0300600 and No. 2016YFA0302104). This work is also supported by the Fund for Shanxi "1331 Project" Key Subjects, China.

\end{document}